\documentclass[12pt]{spieman}  
\usepackage{amsmath,amsfonts,amssymb}
\usepackage{graphicx}
\usepackage{setspace}
\usepackage{tocloft}
\usepackage[version=4]{mhchem}
\usepackage{enumitem}
\usepackage{threeparttable}
\usepackage{hyperref}
\usepackage{multirow}

\DeclareSymbolFont{UPM}{U}{eur}{m}{n}
\DeclareMathSymbol{\umu}{0}{UPM}{"16}
\let\oldumu=\umu
\renewcommand\umu{\ifmmode\oldumu\else\math{\oldumu}\fi}
\newcommand\micro{\umu}

\let\microns \jmicron

\def\msini{M\sin{i}}

\title{Possible Observational Survey Strategies to Maximize Exoplanet Yields: A Report from the Survey Strategies Task Group in the Exoplanet Science Yields Working Group}

\author[a,b,*]{Natasha Latouf}
\author[a,b]{Corey Spohn}
\author[c]{Kevin Fogarty}
\author[d,b]{Christopher C. Stark}
\author[e]{Dmitry Savransky}
\author[a]{Eleonora Alei}
\author[f]{Anthony Boccaletti}
\author[g]{Gagandeep Kaur}
\author[d,b]{Avi M. Mandell}
\author[h]{Rhonda Morgan}
\author[d,b]{Aki Roberge}
\author[i]{Jean-Baptiste Ruffio}
\author[j]{Jean Schneider}
\author[k]{Edward Schwieterman}
\author[l]{Armen Tokadjian}
\author[m]{Daniel Viúdez-Moreiras}

\affil[a]{NASA Postdoctoral Program Fellow, NASA Goddard Space Flight Center, 8800 Greenbelt Road, Greenbelt, MD 20771, USA}
\affil[b]{Sellers Exoplanet Environments Collaboration, 8800 Greenbelt Road, Greenbelt, MD 20771, USA}
\affil[c]{NASA Ames Research Center, Moffett Field, CA, 94035, USA}
\affil[d]{NASA Goddard Space Flight Center, 8800 Greenbelt Road, Greenbelt, MD 20771, USA}
\affil[e]{Sibley School of Mechanical and Aerospace Engineering, Cornell University, Ithaca, NY 14853, USA}
\affil[f]{LIRA, Observatoire de Paris, Université PSL, CNRS, Sorbonne Université, Université Paris Cité, 5 place Jules Janssen, 92195 Meudon, France}
\affil[g]{Technische Universität Graz, Rechbauerstraße 12, 8010 Graz, Austria}
\affil[h]{Jet Propulsion Laboratory, California Institute of Technology, 4800 Oak Grove Dr, Pasadena, CA, 91109, USA}
\affil[i]{Department of Astronomy \& Astrophysics,  University of California, San Diego, La Jolla, CA 92093, USA }
\affil[j]{Paris Observatory, LUX, Meudon, 92190, France}
\affil[k]{Department of Earth and Planetary Sciences, University of California, Riverside, Riverside, USA}
\affil[l]{Jet Propulsion Laboratory, California Institute of Technology, Pasadena, CA 91109, USA}
\affil[m]{Centro de Astrobiología, Spanish National Research Council, 28850
Torrejón de Ardoz, Madrid, Spain}

\cftpagenumbersoff{figure}
\cftpagenumbersoff{table} 
\begin{document} 
\maketitle

\begin{abstract}

We present a variety of possible observational sequences from planet detection to characterization (i.e., survey strategies) for the initial exoEarth detection and characterization survey phase of the Habitable Worlds Observatory. An array of survey strategies were collated by the HWO Exoplanet Survey Strategies Task Group; by investigating and comparing expected resultant exoEarth yields for multiple survey strategies, we can better understand the most efficient strategy to characterize multiple planets in unconventional ways. We present eight varied survey strategies, presented as steps of Detection, Photometry, Orbit Measurement, and/or Characterization, including discussion of possible shortcomings and benefits of each strategy. We present exoEarth yield calculations from the Altruistic Yield Optimizer (AYO) and the Exoplanet Open-Source Imaging Mission Simulator (EXOSIMS) for select strategies, with initial considerations of thermal emission and detector noise. Herein, a yield is the expectation value given the astrophysical assumptions of the number of planets that meet the observational criteria during the survey. We find that AYO and EXOSIMS find the highest yields for \ce{H2O} at 0.9 {\microns}, although the calculated yields vary. When allowing for wavelength optimization within AYO, we find that \ce{H2O} is most optimal to observe at 0.9 {\microns} when given the full range of VIS and NIR values, with an expected exoEarth yield of 29.1. \ce{H2O} and \ce{O2} dual characterization is possible at the expense of yield planet loss, with a calculated possible yield of 20.9 in the VIS. The characterization of \ce{CH4} and \ce{CO2} in the NIR result in the lowest exoEarth yields at 19.9 and 6.4 possible expected planets, respectively. Future work updating AYO and EXOSIMS is necessary to explore further survey strategies that follow different observation procedures, some of which is currently underway. 

\end{abstract}


{\noindent \footnotesize\textbf{*}Natasha Latouf,  \linkable{nlatouf@gmu.edu};\linkable{natasha.m.latouf@nasa.gov} }

\begin{spacing}{2}   

\section{Introduction}
\label{sect:intro}  

The Habitable Worlds Observatory (HWO) will require knowledge of possible scientific productivity well before launch to understand mission requirements. In order to explore scientific productivity, several working groups (WGs) were created, including the Exoplanet Survey Yields sub-Working Group (ESYWG). The ESYWG aimed to understand how requirements on, and the capabilities of, HWO influence exo-Earth yields, and to lay out strategic frameworks to maximize yields. The ESYWG created four task groups to investigate different survey yield aspects; most relevant herein, the Survey Strategies Task Group (SSTG).


A survey strategy is an observational workflow that outlines the sequence of observations leading from planet discovery to characterization. Characterization can indicate anything from detection of a single atmospheric molecule to full biosignature characterization, but in any case refers to searching for atmospheric constituent(s). Once a planet is detected and confirmed, each step of characterization can change, based on the scientific driving question; for instance, after establishing the presence of \ce{H2O} on a potentially habitable planet, is it more efficient to search for \ce{CH4} or \ce{O2}? Or should \ce{H2O} be considered the first step in atmospheric characterization at all? Survey strategies will need to strike a balance between the needs of science (e.g., which biomarkers are the most ``information-rich'', mitigate false positives and negatives) and what is technically achievable. Since our understanding of available biomarkers and technological progress on HWO are both ongoing \cite{gaplist}, what constitutes an optimal survey strategy is necessarily fluid, and may not even be known until after HWO launches. Therefore, a goal of the present effort is to provide tools and a framework for evaluating survey strategies in this evolving landscape.

Survey strategies previously studied for mission concepts predating HWO such as LUVOIR and HabEx \cite{luvoir, habex} follow a ``narrowing" approach as shown in Figure 1-5 of Ref.~\citenum{luvoir}. By setting up a general survey strategy that begins broad and narrows as each step is confirmed, targets for follow-up can be selected as those most likely to be habitable. The current leading strategy is largely focused in the VIS wavelength regime \cite{luvoir, habex}. Recent work by \citenum{livingworldsscdd} shows that these VIS-first strategies are most effective for characterizing Earth analogs with conditions comparable to present-day Earth, but are susceptible to known false-positive scenarios \cite{jktlivingworlds}. These strategies emphasize feasibility with the telescope and instrument technology available at the time, but represent the conservative end of what can potentially be achieved with more advanced tools and techniques. 

Exploration of other survey strategies, such as reversing the order of observations (i.e., beginning with characterization rather than planetary orbit constraint) could illuminate more optimal paths that require less exposure time. Ideally, given a set of criteria for what constitutes characterization (which will require baseline assumptions, such as observations of \ce{H2O} + \ce{O2}), survey strategies will be quantified according to (1) The number of exoplanets predicted to be fully characterized by a survey strategy (yield), where "full characterization" is achieving all of the steps laid out by a given survey strategy, and (2) Whether the strategy can be achieved given planned observatory capabilities. By exploring multiple survey strategies, we can determine the most efficient strategy on a star-by-star and planet-by-planet basis, thus maximizing telescope productivity when launched.  

The number of exoplanets that can be observed at a specified confidence level with a given mission under given telescope and instrument performance assumptions is known as an exoplanet yield. There are currently two well-documented yield estimators for exoplanet direct imaging observations: the Altruistic Yield Optimization (AYO) code \cite{stark19, stark24b} and the Exoplanet Open-Source Imaging Mission Simulator (EXOSIMS) \cite{savransky16, savransky2017exosims}. These tools estimate the yield of the discovery and initial characterization by calculating the exposure times for simulated planets orbiting known stars in a target list based on the mission architecture and observing strategy assumptions, therefore helping guide telescope and mission design trade-offs. The exposure times required to detect key molecular species in an exoEarth’s spectrum can be very long (potentially weeks); thus it is critical that we understand how to achieve our science goals while minimizing exposure times. This can be done by calculating yields for survey strategies that can also guide the future mission design. 

In this work, we present precursor yield results for multiple molecules at varying wavelengths assuming different survey strategies. In $[\S]$ \ref{sect:method} we present the methodology, with descriptions of each yield calculator used. In $[\S]$ \ref{sect:surveystrategies} we outline the benefits and drawbacks to each survey strategy proposed, with the yield results for few of the proposed survey strategies presented in $[\S]$ \ref{sect:results}. In $[\S]$ \ref{sec:discuss} we discuss the presented results, analyze the impact for future calculations, and outline future work. In $[\S]$ \ref{sec:conc} we present our conclusions.

\section{Methodology}
\label{sect:method}

\subsection{Altruistic Yield Optimizer}

AYO optimizes observation strategy on multiple axes, including exposure times and selected targets, to calculate the maximum yield possible of any given survey, in units of planets per survey. AYO calculates the exposure time per planet given an instrument model and background noise, for $\simeq10^{5}$ synthetic user-defined planets around each star (e.g., ``exoEarth candidates"). The planets are sorted by exposure time, and their completeness as a function of time \cite{brown2004} is calculated. The derivative of completeness, dC/dt, is also calculated, and the optimal value of dC/dt is determined for all observations, thus optimizing the targets, visits per star, and exposure time per visit for the maximum resultant yield. AYO can also optimize the bandpass (i.e. wavelength for observation) chosen per star. The input target list used is the HWO Preliminary Input Catalog (HPIC) \cite{tuchow24}, which contains $\sim$13,000 stars within 50 pc complete to 12$^{\rm th}$ TESS magnitude. Further information on AYO and HPIC can be found in Refs.~\citenum{stark19, tuchow24}. In this work, we will be utilizing version 17 of AYO implemented into the Framework for Remote Implementation of Demand-based Altruistic Yields (FRIDAY); updates available in v17 are described in Ref.\citenum{latoufresolving}. All yields have intrinsic uncertainty, that is less than the uncertainty due to other effects. Other works have calculated yield uncertainties\cite{savransky10}, which can be used to benchmark this work.

The high-level astrophysical assumptions are mirrored to Table 1 in \citenum{stark24b}. For our baseline mission, we adopt parameters similar to those for LUVOIR-B adopted by \citenum{stark24a}. We adopt a segmented, off-axis primary mirror with an inscribed diameter of 6.7 m. We assume two parallel detection channels in the VIS and NIR. We investigate the effects of thermal emission by calculating yields with an effective thermal emission of 290K, as well as zero K (the second value is used only to eliminate the impact of thermal self-emission from the results for comparison). The characterization wavelength, as well as the detection wavelength, is free to be optimized on a star-by-star basis. We adopt the coronagraph used in the LUVOIR-B study, a deformable mirror-assisted charge 6 vortex coronagraph (DMVC).  We adopt the noise floor (i.e. $\Delta mag_{floor}$) used by \citenum{stark25arxiv}, and assume a moderate detector noise scenario, described in Table~\ref{tab:missionparams}. We adopt a raw contrast floor of $1\times10^{-10}$ and a minimum working angle of 1$\lambda/D$. In order to provide more optimistic performance metrics, we assume generic detector parameters resembling a Skipper CCD \cite{bebek}, which is not the same as the initial LUVOIR report, as it has a higher QE near 1 {\microns} than, for example, the Roman EMCCD \cite{teledyne, romanemccd}.

As we are only assuming spectroscopic observations for molecular characterization, certain parameters must be adapted. We assume that an integral field spectrograph (IFS) will be the dispersing instrument, which carries additional optics resulting in a 30\% reduction in throughput. We also adopt 6 pixels per spectral bin per lenslet at the Nyquist wavelength of 500 nm. Additionally, we mandate six initial visits to every target following the LUVOIR study \cite{luvoir}, with the assumption that this will result in at least three detections of a planet; three detections are assumed necessary for confirmation and orbit determination. The signal-to-noise ratios (SNRs) given to AYO are produced from band-limited spectral retrievals using the Bayesian Analysis for Remote Biosignature Identification on exoEarth (BARBIE) methodology\cite{latouf23,latouf24a,barbie3}. The results from BARBIE assume varying molecular abundances (indicated per section) and a 50\% cloudiness fraction. Table~\ref{tab:missionparams} provides a summary of all mission parameters given to AYO, and Table~\ref{tab:snr_requirements} provides the input SNRs per wavelength.

We calculate yields across the VIS and NIR wavelength regimes - however thermal emission and detector noise become much more pervasive in the NIR than the VIS. In order to understand which component is more impactful on total yields, we conducted the initial \ce{H2O} yields with each combination of thermal emission, no thermal emission, noise, and no noise. We assume that there is not a single channel that can extend across both the VIS and NIR wavelengths, therefore a detector transition has to occur in order to observe at all wavelengths. We assume this detector transition occurs at 1.0 {\microns} (i.e. at wavelengths shorter than 1.0 {\microns}, it is the VIS detector and at wavelengths longer than 1.0 {\microns} it is a separate, NIR detector). To account for the switch in detector, there are changes to the noise and resolving power that occur at 1.0 {\microns}. Longer than 1.0 {\microns}, all resolving powers are 70 rather than 140 as in the VIS. Additionally, the number of pixels will scale as 1/0.5 in the NIR, thus there will be 4x as many pixels as in the VIS. In order to account for this effect, all noise values are divided by 4 in the NIR in order to not overestimate the noise. 

\begin{table}[h!]
\centering
\resizebox{\columnwidth}{!}{%
\begin{threeparttable}
\caption{Coronagraph Mission Parameters}
\begin{tabular}{ccc}
            \hline
            \hline
            \textbf{Parameter} & \textbf{Value} & \textbf{Description}\\
            \hline
             & & \textbf{General Parameters}\\
            $\sum_\tau$ & 2 years & Total exoplanet science time out of an assumed 5 year mission\\
            $\tau_{dynamic}$ & 1.1 & Multiplier overhead to touch up dark hole\\
            $\tau_{static}$ & 2.32 hrs & Overhead for slew, settling, and digging dark hole\\
            $X$ & optimized per planet & Inscribed photometric aperture radius to extract planet signal ($\lambda/D_{LS}$)\\ 
            $\varsigma_{floor}$ & $1\times10^{-10}$ & Raw contrast floor \\
            Post-Processed Noise Floor Contrast & $3\times10^{-12}$ & Setting the 1$\sigma$ noise floor contrast (uniform over FOV and independent of raw contrast) \\
            $T_{contam}$ & 0.95 & Effective throughput as driven by contamination \\
            $T$ & 0, 290$^{a}$ & Temperature in Kelvin \\
            \hline
             & & \textbf{Detection Parameters}\\
            $\lambda_{d}$ & 650 $nm^{b}$ & Central wavelength for detection \\
            $SNR_{d}$ & 7 & Required SNR for detection \\
            $T_{optical}$ & 0.338$^{b}$ & End-to-end reflectivity at $\lambda_{d}$ \\
            $\tau_{d,limit}$ & 2 months & Detection time limit, overheads included \\
            \hline
             & & \textbf{Characterization Parameters}\\
            $\lambda_{c}$ & 750 $nm^{b}$ & Central wavelength for characterization \\
            $SNR_{c}$ & [5 - 20]$^{c}$ & SNR required for characterization with IFS \\
            $R$ & 70, 140 & Spectral resolving power for IFS in NIR, VIS \\
            $T_{optical,IFS}$ & 0.233 & End-to-end reflectivity at $\lambda_{c}$ for IFS\\
            $\tau_{c,limit}$ & 2 months & Characterization time limit, overheads included \\
            \hline
             & & \textbf{Detector Parameters}\\
            $n_{pix,d}$ & 1 & Number of pixels in photometric aperture per imager at $\lambda_{d,\#}$\\
            $n_{pix,c}$ & 6 & Number of pixels per spectral bin in coronagraph IFS at $\lambda_{c}$ \\ 
            $\xi$ & $5\times10^{-5}$$ e^{-} pix^{-1} s^{-1}$ & Dark current\\
            RN & $0$$ e^{-} pix^{-1} s^{-1}$ & Read noise\\
            $\tau_{read}$ & 1000 s & Time between reads\\
            CIC & $1.3\times10^{-3}$$ e^{-} pix^{-1} s^{-1}$ & Clock induced charge\\
            $T_{QE}$ & 0.9 & Raw detector QE at all wavelengths \\
            $T_{dQE}$ & 1 & Effective throughput due to bad pixel/cosmic ray mitigation \\
            $T_{core}$ & 0.75 & Core Coronagraph Throughput \\ 
            \hline
            \hline
\end{tabular}

\begin{tablenotes}
\item $^{a}$Values for temperature that result in effectively no thermal emission, and full thermal emission.
\item $^{b}$Values for the most likely bandpass, however AYO optimizes bandpass and will adjust.
\hspace{0.35cm} \item $^{c}$SNRs for strong \ce{H2O}, and dual \ce{H2O}/\ce{O2} detection. See \ref{tab:snr_requirements} for further information. 
\end{tablenotes}
\label{tab:missionparams}
\end{threeparttable}}
\end{table}

\subsection{EXOSIMS}

To validate the yields produced by AYO and investigate the impact of dynamic scheduling constraints on survey efficiency, we utilized  EXOSIMS\cite{savransky16}. Unlike AYO, which calculates exposure times based on a
probabilistic view of the average HWO mission, EXOSIMS simulates the full timeline of a mission, ensuring the observations do not overlap and managing the scheduling competition for time between new detections and characterizing known planets. The output unit is expected number of characterized planets per survey, specifically exoEarths in this simulation.

For these simulations, we translated the inputs used in the AYO analysis (see Table~\ref{tab:missionparams}) directly into EXOSIMS configuration files using \texttt{yieldplotlib} \cite{Spohn2025}. We generated a synthetic universe of planets based on the ``Nominal'' occurrence rates shown in Figure 17 of Ref.~\citenum{dulzJointRadial2020}, following the LUVOIR and HabEx reports, and simulated a 5-year mission with 40\% of the observing time allocated to the exoEarth survey.

We used the \texttt{orbix} dynamic scheduler, a probability-driven algorithm detailed in Ref.~\citenum{spohn2026}. This scheduler optimizes the transition from blind search detection to targeted follow-up by tracking the orbital uncertainty of each candidate. It forecasts the probability of detection ($P_{det}$) for future epochs and schedules characterization visits only when the $P_{det}$ exceeds a strict threshold (0.95), thereby minimizing wasted integration time on planets obscured by the IWA or their host star. To match AYO's assumption that characterizations occur at the optimal time, we assume full orbital knowledge after the initial detection.

\begin{table}[h!]
\centering
\resizebox{0.8\columnwidth}{!}{%
\begin{threeparttable}
\caption{Wavelength-Dependent SNR Requirements (20\% Bandwidth)}
\label{tab:snr_requirements}
\begin{tabular}{cc|c}
\hline
\hline
\textbf{Wavelength ($\mu$m)} & \textbf{Required SNR} & \textbf{Scenario Applicability} \\
\hline
0.720 & 17 & \multirow{2}{*}{\ce{H2O} Only} \\
0.736 & 17 & \\
\hline
0.752 & 15 & \multirow{11}{*}{\ce{H2O} \& \ce{O2}} \\
0.768 & 14 & \\
0.784 & 14 & \\
0.801 & 14 & \\
0.819 & 13 & \\
0.830 & 12 & \\
0.848 & 12 & \\
0.867 & 12 & \\
0.879 & 12 & \\
0.898 & 12 & \\
0.911 & 10 & \\
\hline
0.930 & 9 & \multirow{5}{*}{\ce{H2O} Only} \\
0.944 & 6 & \\
0.957 & 6 & \\
0.978 & 6 & \\
0.992 & 5 & \\
\hline
\hline
\end{tabular}
\begin{tablenotes}
\small
\item \textbf{Note:} EXOSIMS mission parameters (telescope diameter, throughput, etc.) are identical to the AYO parameters listed in Table~\ref{tab:missionparams}.
\end{tablenotes}
\end{threeparttable}}
\end{table}

\section{Survey Strategies}
\label{sect:surveystrategies}

\subsection{Strategy Classification}

\begin{figure*}
    \centering
    \includegraphics[width=0.9\linewidth]{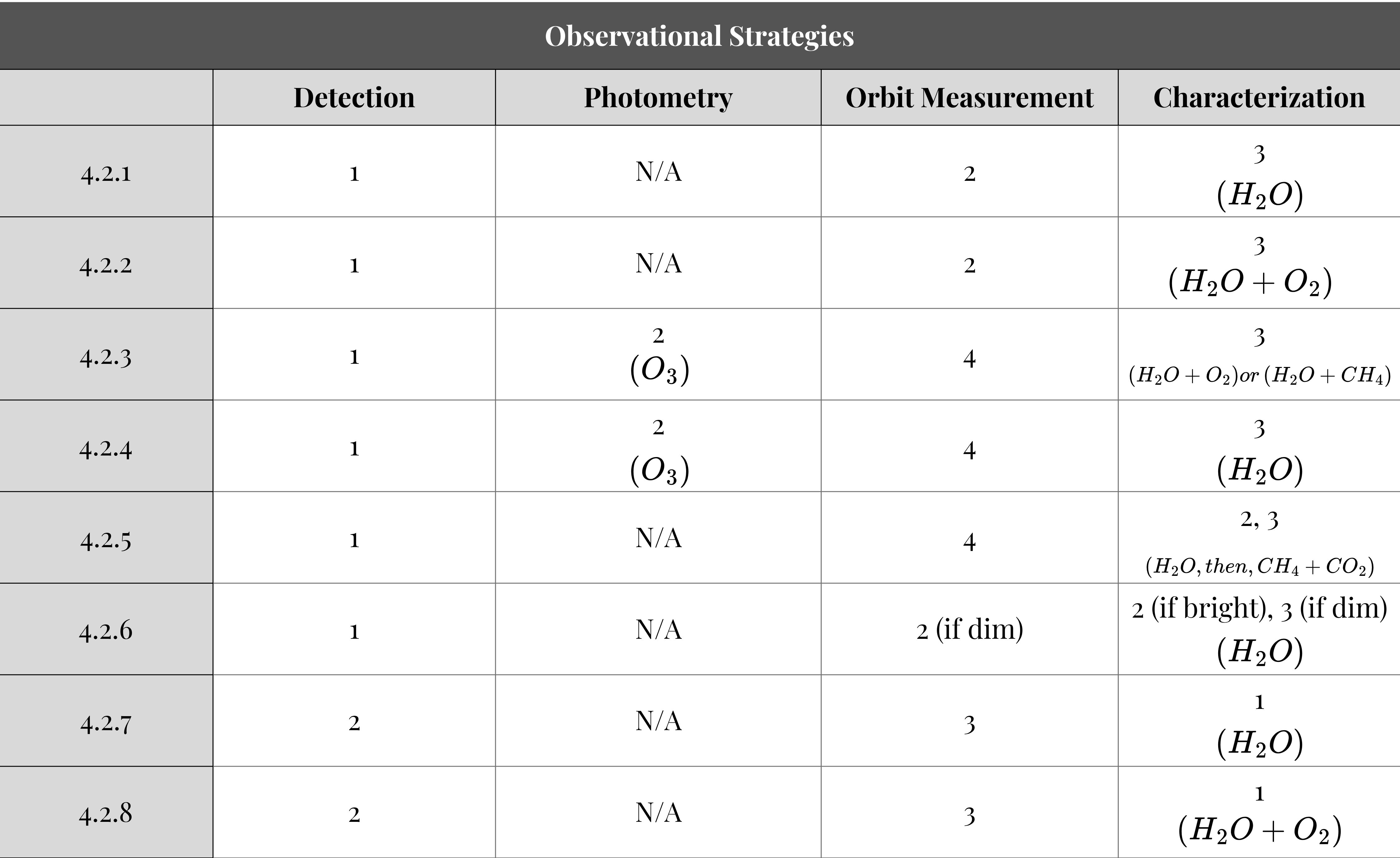}
    \caption{A table showing the overarching order of operations for each observational strategy. The rows are the corresponding subsections, and the columns are the survey strategy components. Each number indicates the placement of the survey strategy component in the observational scheme, and the molecules for characterization are indicated in each box. For further information, please see the full descriptions in Section~\ref{sect:strategies}. *N/A indicates that the observational method is not used for the survey strategy.}
    \label{fig:tablesummary}
\end{figure*}

This list of eight survey strategies is not comprehensive of possible HWO survey strategies, but is a set of initial concepts put forward by SSTG members and considered to be potentially favorable strategies. Each strategy is broken down into the order of observations, the driving logic, and any prior works that are key to the strategy were summarized. An overarching summary of all observational strategies is presented in Figure~\ref{fig:tablesummary}, indicating the subsection of each strategy and the numbered order of operations.

All of the strategies incorporate coronagraphic direct imaging and coronagraphic direct spectroscopy; however, they do not include the impact of precursor, complementary, or second-generation observations (e.g. ground-based RV, space-based astrometry, starshade, etc). No assumptions are made as to available filters and spectral resolutions. As a baseline for future work, we consider the capabilities and instrument suite of LUVOIR-B detailed in Ref.~\citenum{luvoir}. We classify survey strategies according to what the strategy intends to observe (i.e., what observables constitute detection of a candidate exo-Earth and what constitute characterization) and how we want to observe it and in what order.

Components of a survey strategy generally include:
\begin{enumerate}
    \item \textbf{Detection}: photometric (i.e. imaging) or spectroscopic observation that provide the initial detection of exoEarth candidates (EECs). As stated above, for the purposes of this work we do not assume any precursor observations that resulted in a detection, and thus we define an EEC as any planet we can detect with HWO coronagraphic imaging that is approximately Earth-sized in the habitable zone of its host star. 
    \item \textbf{Photometry}: photometric observation of candidate habitable planets, with either the central wavelength and bandwidth or filter name specified. This is assumed to be in filters or spectral channels other than the detection filter, and does not utilize spectroscopy. No assumptions about the technical feasibility of any specific bandwidth were applied, and we assume a constant 20\% filter bandpass.
    \item \textbf{Orbit Measurement}: Series of observations to constrain the planet's orbit, with the observing cadence and wavelength band provided.
    \item \textbf{Characterization}: Spectroscopic biomarker observation, with only the desired molecule(s) and their wavelengths specified. 
\end{enumerate}

Some modes can be combined as needed, in order to achieve multiple goals with the same observations. For example, an observation with tags Orbit Measurement and Photometry consists of a set cadence of observations utilizing both parallel photometry channels. Additionally, we note that beyond the initial characterization steps included in each survey strategy, there are many additional possible observations that are not included, such as studying surrounding objects (i.e., rings and/or moons), and further analysis of the planetary surface (e.g., vegetation red edge and/or oceans) that would only be performed on a strong candidate for extra deep follow-up after each step in the presented survey strategies is completed. We anticipate that HWO will be able to measure or constrain some of these planetary characteristics on a case-by-case basis if we discover feasible targets.

\subsection{Strategies}
\label{sect:strategies}

\subsubsection{Detection $\rightarrow$ Orbit Measurement $\rightarrow$ Characterization of \ce{H2O}}
\label{sect:h2ofirst}

In the initial LUVOIR final study, the observational method by which to detect and characterize a planet was influenced by the perceived level of difficulty for each step (i.e. the potential wasted observational time), with the aim to efficiently establish that a planet is an exoEarth candidate (EEC). First, the planet must be discovered, which was perceived as the easiest step and the least potential waste of observational time. This is assumed to occur with photometric observations at 0.5 {\microns}, as the most optimal location for minimal noise and most efficient Nyquist sampling\cite{luvoir}. 

The next most difficult step, thus requiring more observational time, was to re-observe the target with multiple epochs to determine the orbit, and thus determine if the planet was within the habitable zone (HZ) of the host star. Ref.~\citenum{horning19} found that three equally spaced epochs over at least one half of an orbital period is the most efficient observational method to minimize the uncertainty in the semi-major axis and eccentricity to below 10\% for an observed planet. While three epochs is sufficient for systems with an inclination of $\leq$80$^{\circ}$, four epochs are needed for highly-inclined systems\cite{blunt17, habex}. Since exoEarths would certainly require longer exposure times for characterization, determining the orbit first to ensure the planet is within the HZ may avoid costly spectral observations on non-habitable planets. For the AYO simulations, we do not assume three equally spaced epochs, but instead space the observations based on what maximizes completeness in the shortest amount of time, with 6 epochs, not all of which result in a detection.

The most difficult perceived step, and the largest utilization of observational time, was to search for signs of habitability through the detection of water vapor (\ce{H2O}) in the planetary atmosphere. After establishing a planet's orbit lies in the HZ, spectroscopic observations are utilized to attempt to detect \ce{H2O} to determine the potential habitability of the planet. \ce{H2O} can also be detected photometrically\cite{latoufresolving}, depending on the abundance and achieved signal-to-noise ratio (SNR), however no other molecules in the VIS or NIR can be sufficiently detected with photometry, thus we assume that characterization of a molecule occurs spectroscopically unless otherwise specified. Depending on the abundance of \ce{H2O} present in the planetary atmosphere, additional observational time may be required to sufficiently constrain the abundance or lower limit. Whether further spectroscopic observations are necessary to fully characterize \ce{H2O} will depend on the detection of the planet, and if there is indication that it is an EEC; this will vary from star-to-star.

These three steps (detection, orbit measurement, and molecular characterization), we will refer to as the ``initial discovery survey", although the molecule selected for characterization will vary from survey to survey. It is assumed that all of the steps take place in the two year time span allotted for exoplanet science, and all reported yields are for a 2-year survey time. Henceforth, we shall reference the initial discovery survey and describe any differences in subsequent survey strategies. 

\subsubsection{Detection $\rightarrow$ Orbit Measurement $\rightarrow$ Characterization of \ce{H2O} and \ce{O2} simultaneously}
\label{sect:o2h2o}

In this strategy, we alter the initial discovery survey slightly. Detection and orbit measurement remain the first two steps as described in Section~\ref{sect:h2ofirst}, however \ce{H2O} and \ce{O2} are characterized simultaneously with spectroscopy rather than \ce{H2O} individually. This will require observation at shorter wavelengths in the VIS as the \ce{O2} absorption band is centered at 0.76 {\microns} and can be detected with 20\% bandpasses up to 0.83 {\microns} as found in Ref.~\citenum{latouf24a}. Per Ref.~\citenum{latouf24a}, dual detection of \ce{H2O} and \ce{O2} is possible for modern Earth analogs at a higher SNR than is required to only detect \ce{H2O} at 0.9 {\microns}. Requiring a higher SNR indicates that a longer exposure time is necessary to achieve the desired detections. Investigation is necessary to determine whether it is more efficient to seek a yes/no observation of \ce{H2O} and \ce{O2} and re-observe to characterize the molecules, or to immediately seek robust observation to measure abundances of both molecules simultaneously. This will determine whether to seek this strategy rather than search for \ce{H2O} individually, i.e., if there is already a robust observation planned (such as for a known target of interest), it is more efficient to seek \ce{H2O} and \ce{O2} simultaneously than \ce{H2O} individually. 

\subsubsection{Detection $\rightarrow$ Characterization of \ce{H2O} $\rightarrow$ Characterization of \ce{CO2}/\ce{CH4} at the red edge $\rightarrow$ Orbit Measurement}
\label{sect:rededgeafterh2o}

This strategy deviates from the initial discovery survey in multiple ways. It postulates that conducting a survey long enough to get sufficient spectral information to characterize a molecule initially could be more efficient than following the initial discovery survey. The first step of detection remains, however the second step is no longer orbit measurement. As a result, fewer epochs are required initially to detect the planet as the orbit is not immediately measured. Instead, after detection, \ce{H2O} is characterized followed by the characterization of \ce{CO2} and/or \ce{CH4}, i.e. spectroscopic observations are immediately taken. The strategy ends with the orbit measurement, however the strategy to characterize multiple molecules prior to determining the orbit would involve multiple observations that could occur at three or four different epochs, which will then suffice to determine the orbit for any system without additional photometric observations. Thus, by characterizing multiple molecules, the orbit will be determined as the planet is characterized (assuming simultaneous imaging observations). This strategy will determine how many planets have the potential for multi-wavelength characterization through the NIR. However, as atmospheres change over time, varying epochs may result in non-consistent molecular abundances.

\subsubsection{Detection $\rightarrow$ Photometry of \ce{O3} $\rightarrow$ Characterization of \ce{H2O} and \ce{O2} OR \ce{H2O} and \ce{CH4} $\rightarrow$ Orbit Measurement}
\label{sect:complex}

\begin{figure*}
    \centering
    \includegraphics[width=0.5\linewidth]{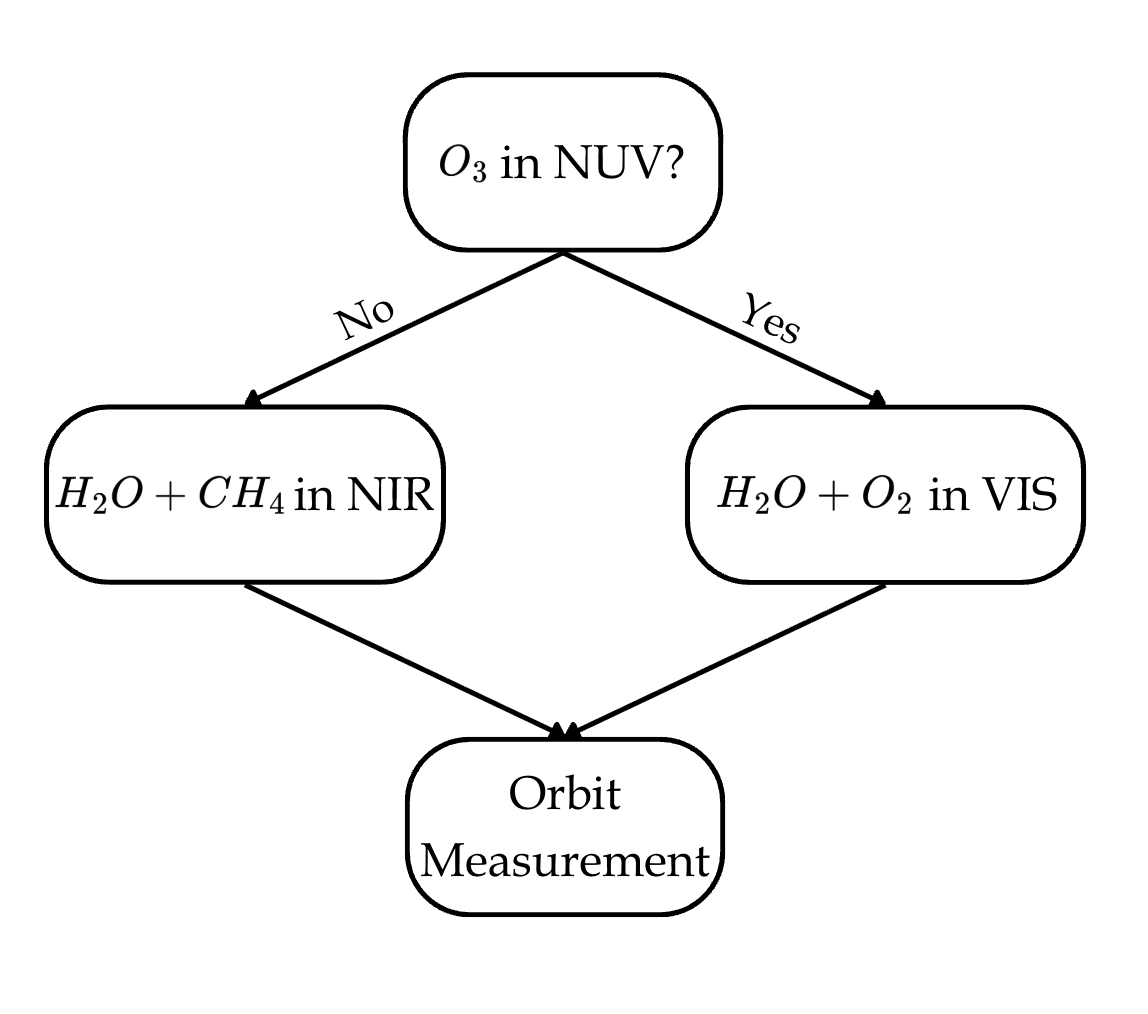}
    \caption{Flowchart of the observational strategy suggested in Section~\ref{sect:complex}.}
    \label{fig:flowchart}
\end{figure*}

The initial survey strategy is significantly altered in this proposed survey strategy. Due to the complexity of this strategy, a flowchart has been provided in Figure~\ref{fig:flowchart} to illustrate the process. Although still beginning with detection, the next step is to attempt to characterize \ce{O3} in the NUV using photometry. This will require a minimum of three photometric points\cite{livingworldsscdd}, and would signify an oxygen-rich atmosphere if found. If \ce{O3} is detected, spectroscopic observations to characterize \ce{H2O} and \ce{O2} in the VIS will occur. If \ce{O3} is not detected, spectroscopic observations to characterize \ce{H2O} and \ce{CH4} will be conducted in the NIR. Following the molecular characterization decision, two more epochs will be necessary to determine the orbit, as only one would have been conducted by this stage. 

While NUV coronagraphy has yet to be demonstrated on-sky, it appears progressively more promising as HWO development continues, and could improve our ability to detect oxygen-dominated atmospheres for planets at small inner working angles (IWAs). Additionally, the \ce{O3} feature in the NUV is intrinsically strong, leading to a very deep absorption feature with small abundances, and increasing detectability of oxygen-rich atmospheres at most abundances through Earth-time \cite{livingworldsscdd}. Thus, characterizing \ce{O3} should be an achievable endeavor if oxygen is present in the atmosphere; however, the development of a UV high-contrast coronagraph poses technical difficulties to this observation. \ce{O3} is highly dependent on \ce{H2O} photolysis products, thus \ce{H2O} characterization will be critical to providing context and is thus included in either following step. However, as \ce{O3} and \ce{O2} are also intrinsically tied\cite{domagalgoldman14,meadows17}, we will likely be able to infer the \ce{O2} abundance -- i.e. if Proterozoic levels of \ce{O3} are characterized, that will imply undetectable \ce{O2} abundances\cite{latouf24a}, and thus can move directly to characterization of \ce{H2O} and \ce{CH4}. Carefully selecting the bandpasses for spectroscopy so that multiple molecules are detectable can increase the probability of at least one molecular detection. \ce{O3} detection using a starshade was the focus of the probe-class mission concept Occulting Ozone Observatory (O3) mission \cite{savranskyo3, o3mission}, highlighting how the critical the detection of \ce{O3} was considered well before HWO. 

\subsubsection{Detection $\rightarrow$ Photometry of \ce{O3} $\rightarrow$ Characterization of \ce{H2O} $\rightarrow$ Orbit Measurement}
\label{sect:o3first}

This strategy is similar to that outlined in Section~\ref{sect:complex}, with simplifications. Again we begin with detection, and photometric characterization of \ce{O3} as outlined above. However, following a either \ce{O3} detection or non-detection, the characterization of \ce{H2O} remains the next observational step. Orbit measurement is again the final step, requiring two additional epochs to sufficiently determine the orbit. By removing the characterization of \ce{O2}, the resolving power requirements for the VIS detector become far less stringent. \ce{O2} is a narrow feature that requires a high R for optimal detection, where both \ce{O3} in the NUV and \ce{H2O} at 0.9 {\microns} are possible to detect and characterize photometrically at low R as found in Ref.~\citenum{livingworldsscdd,latoufresolving}, and thus require lower SNRs.

\subsubsection{Detection $\rightarrow$ Characterization of \ce{H2O} for bright planets, otherwise Orbit Measurement $\rightarrow$ Characterization of \ce{H2O}}

This strategy is the same as the initial survey strategy, unless the planet is bright and preferably nearby. After detection, if the planet is bright, the second step will be the characterization of \ce{H2O}. Characterizing \ce{H2O} is desirable for any type of exoplanet, regardless of presence in the HZ, and thus the orbit determination can move to the final step with an additional two epochs necessary after the characterization of \ce{H2O}. 

\subsubsection{Characterization of \ce{H2O} $\rightarrow$ Detection $\rightarrow$ Orbit Measurement}
\label{sect:h2odetorb}
This strategy inverts the initial discovery survey, while all the same steps remain present. Herein, it is again thought that conducting a survey long enough to gain sufficient spectral information to characterize \ce{H2O} while simultaneously detecting the planet could be more efficient. This requires a longer initial investment of exposure time per target, and is then ended with the orbit determination if and only if \ce{H2O} is detected, which will require multi-epoch observations to determine the orbits. This strategy is only applicable to the subset of targets that are expected to be the most easily observable. However, by tuning the observations to immediately be sensitive to \ce{H2O}, we can re-observe to identify the orbit, habitable zone placement, etc. with the knowledge of a strong potentially habitable planet candidate. Certain coronagraph types (e.g. Phase-apodized-pupil Lyot Coronagraphs (PAPLC)) are excellent for characterization due to the chromaticity of the coronagraph, which results in better contrast in the narrow band (i.e. characterizations) but take longer to detect with the broad band (i.e. detection and orbit measurement); thus, this strategy hinges on differences between hardware solutions and the resulting exoplanet yields per technology type. 

\subsubsection{Characterization of \ce{H2O} and \ce{O2} $\rightarrow$ Detection $\rightarrow$ Orbit Measurement}

This strategy is essentially identical to Section~\ref{sect:h2odetorb}, but rather than characterizing \ce{H2O} alone, characterizing \ce{H2O} and \ce{O2} simultaneously will be the first step. Any targets that result in a detection of \ce{H2O} and/or \ce{O2} will be re-observed with multiple epochs to determine the orbit. This strategy would require a higher initial SNR to dual detect \ce{H2O} and \ce{O2} as a first step, and thus require higher exposure time. Bandpass selection will also be a critical factor.

\section{Results}
\label{sect:results}

The survey strategies outlined in Section~\ref{sect:strategies} are complex, and many require adaptations to current software to fully explore the efficiencies to be gained. However, a key element to every strategy is the selection of the most optimal bandpass or wavelength for molecular characterization. Each molecule's absorption peaks at different wavelengths, with some presenting features at multiple wavelengths. We present precursor work studying the most optimal wavelength for observation per molecule, in order to narrow the parameter space for future works to fully study each survey strategy. In our yield calculations, we assume that each strategy is that of the initial discovery survey, and vary the molecule for characterization. This will constrain the best wavelength for observation per molecule, regardless of the step in an observational framework. We first begin by presenting the normalized yields for AYO and EXOSIMS over a variety of wavelengths, and following that investigation, present AYO results wherein the wavelength is optimized. 

\subsection{Normalized Yields Over Wavelength}

\begin{figure*}
    \centering
    \includegraphics[width=0.9\linewidth]{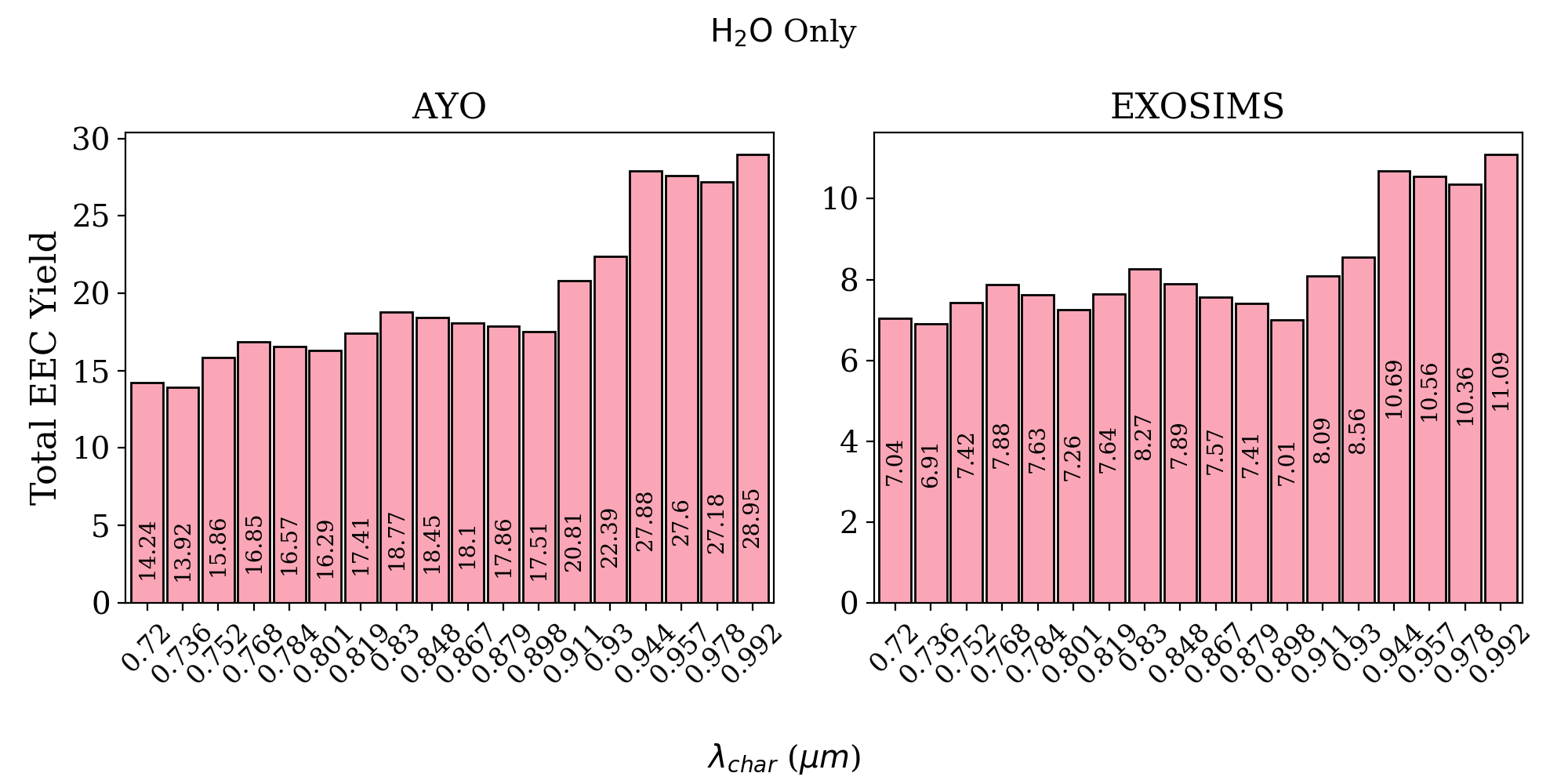}
    \caption{Yields for the detection and characterization of \ce{H2O} in the VIS. X-axis: calculated yield. Y-axis: characterization wavelength in $\mu$m. The total calculated yield per wavelength is printed on each bin. \textit{Left Panel:} AYO yields. \textit{Right Panel:} EXOSIMS yields. \textbf{AYO and EXOSIMS both prefer the same wavelength for characterization of \ce{H2O}.}}
    \label{fig:h2o_ayo_exosims}
\end{figure*}

AYO and EXOSIMS have many differences in their calculations of total exoEarth yields. While other works have aimed to benchmark these and other yield exposure time calculators \cite{starkbenchmark}, we leave further investigation of these differences and their causes to future works. Figure~\ref{fig:h2o_ayo_exosims} presents the calculated yields for each characterization wavelength option, calculated with AYO (left panel) and EXOSIMS (right panel). The total EEC yield is shown on the y-axis, with the calculated value per wavelength printed per bin, with wavelength on the x-axis. Although the total calculated yields vary significantly between the two panels, both find that a long wavelength edge of 0.992 {\microns} for characterization of \ce{H2O} results in the highest calculated yields. Additionally, the calculated yields across both panels are consistent in relative trends across wavelength, with the 0.9 {\microns} feature resulting in the highest yields from both AYO and EXOSIMS.

\begin{figure*}
    \centering
    \includegraphics[width=0.9\linewidth]{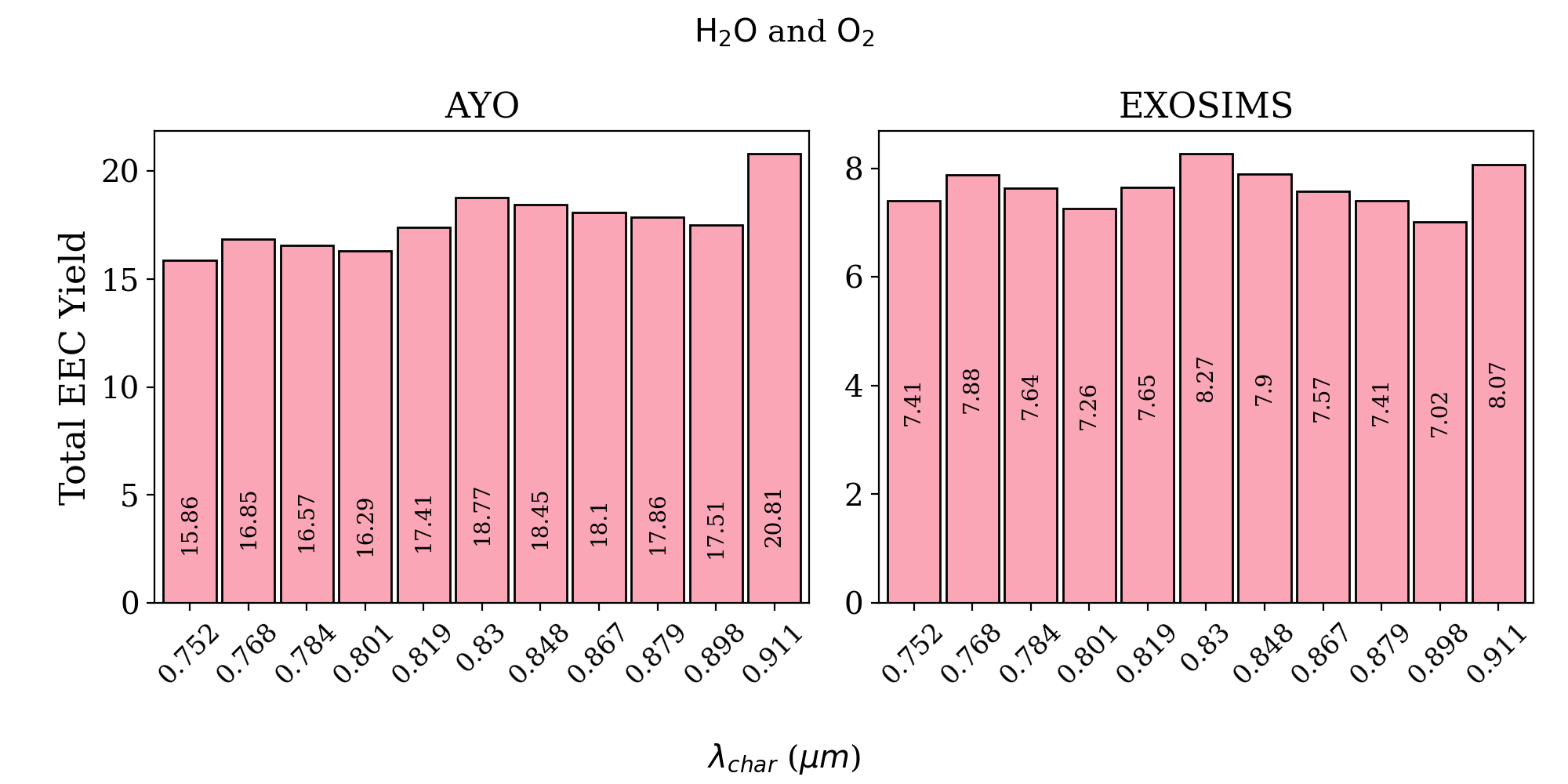}
    \caption{Yields for the detection and characterization of \ce{H2O} and \ce{O2} in the VIS. All other plot facets are identical to Figure~\ref{fig:h2o_ayo_exosims}. \textbf{The preferred wavelength for dual characterization is inconclusive through EXOSIMS, but strongly found with AYO.}}
    \label{fig:h2o_o2_ayo_exosims}
\end{figure*}

In Figure~\ref{fig:h2o_o2_ayo_exosims}, we present the calculated yields for simultaneous \ce{H2O} and \ce{O2} characterization for AYO (left panel) and EXOSIMS (right panel), with all other plot facets identical to Figure~\ref{fig:h2o_ayo_exosims}. When comparing the yields, we find that the two highest yields are calculated at the same wavelengths between AYO and EXOSIMS (0.83 {\microns} and 0.911 {\microns}). AYO calculates the highest yields at a long wavelength edge of 0.911 {\microns}, and EXOSIMS calculates the highest yields at 0.83 {\microns}. All calculated EXOSIMS yields are within 1.1 and fall within an average standard deviation of 2.51. However, while calculating the yields per wavelength will indicate the preferred wavelength for maximum yields, the most optimal wavelength for observation will vary on a target-to-target basis.

\subsection{Wavelength Optimization}

AYO and EXOSIMS both display a curve of normalized yields that prefer the same wavelengths for observations of \ce{H2O} and \ce{H2O} with \ce{O2}. However, these results are constrained to the visible wavelengths, with each yield calculated at an individual wavelength. To investigate the most optimal wavelength for observation per molecule, below we present AYO results wherein an array of wavelengths are given and permitted to be selected preferentially.

\subsubsection{Characterization of \ce{H2O}}

Prior works\cite{feng18, latouf23, stark24a} have investigated the detectability of \ce{H2O} in the VIS. Here, we extend our study into the NIR, which has so far not been examined. To do so, we run yield studies using an updated version of AYO (described in Section~\ref{sect:method} above). As a first step, we reproduced the results found in \citenum{stark24a} and found nearly identical results, with our calculated yield of 28.9 compared to 28.2 in the scenario G explored in \citenum{stark24a} (see Table 1 in \citenum{stark24a}). There are minor differences due to the change in calculation methods in varying versions of AYO (v17 in this work vs v12 in \citenum{stark24a}) 

\begin{figure*}
    \centering
    \includegraphics[width=0.75\linewidth]{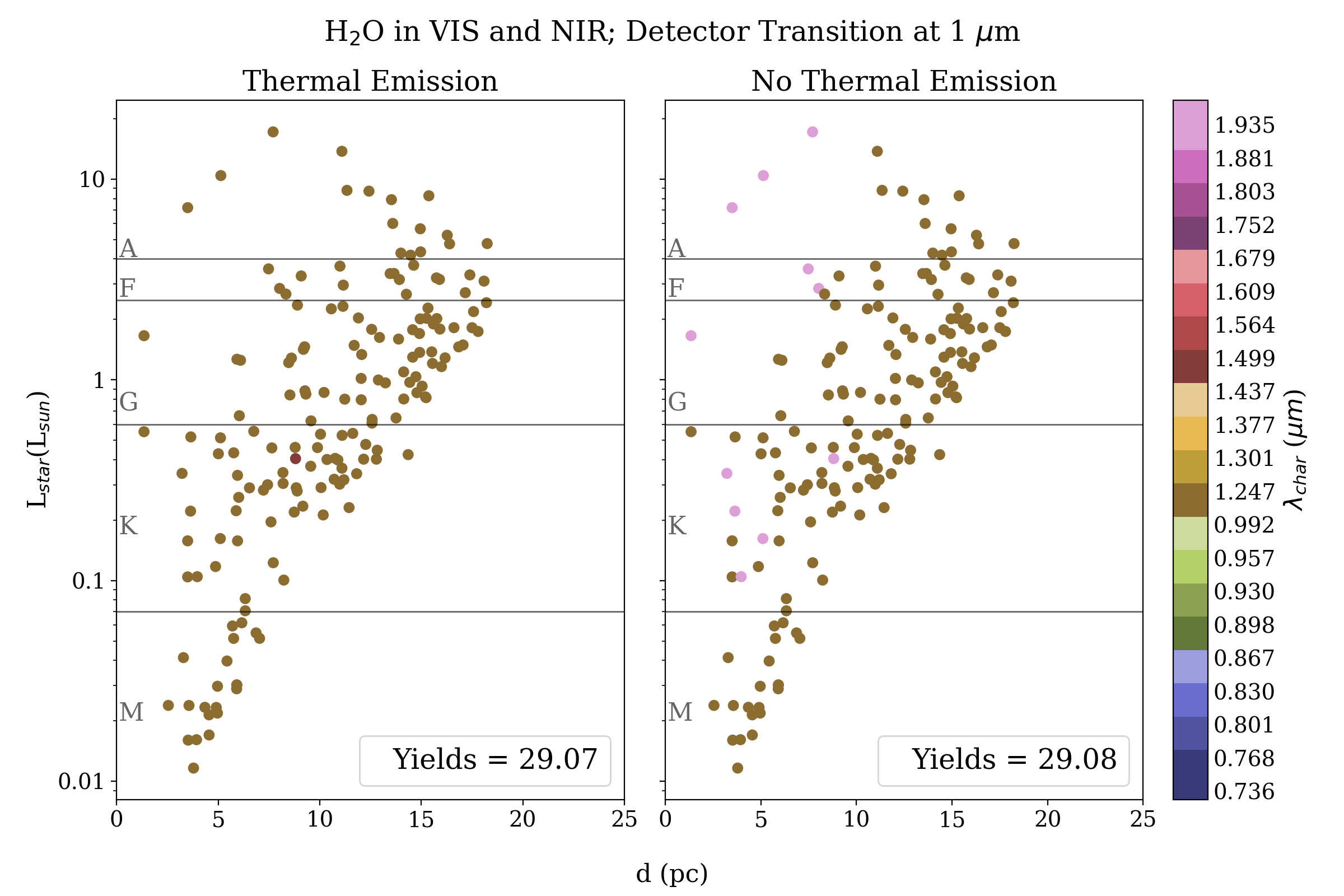}
    \caption{Yields for the detection and characterization of \ce{H2O}. \textbf{Y}-axis: luminosity in L$_{star}$/L$_{sun}$. \textbf{X}-axis: distance in parsecs. The color scale is the most optimal wavelength at which to characterize for the given molecule, with the color bar indicating the long wavelength edge of the bandpass. The luminosity regions for an A, F, G, K, and M star are shown in grey lines. Noise is included in both scenarios. \textit{Left Panel:} full thermal emission. \textit{Right Panel:} no thermal emission. \textbf{Wavelengths shorter in the NIR are preferred for the detection and characterization of \ce{H2O}}}
    \label{fig:ayo_h2o_vis_nir_1000_temp}
\end{figure*}

Initially, we aim to begin to quantify the effects of thermal emission and detector noise. We present the results across the VIS and NIR with and without thermal emission in Figure~\ref{fig:ayo_h2o_vis_nir_1000_temp}. Detector noise is included in both cases. Each plot shows the selected targets as functions of luminosity and distance, with the long wavelength edge of the bandpass as the color scale. For the remainder of the work, we will clarify the corresponding bandpass center. While we find the calculated yields are nearly identical, the wavelength selected for characterization varies. In the left panel with thermal emission present, we find that a long wavelength edge of 0.992 {\microns} (i.e. a bandpass centered on 0.9 {\microns}) is preferred for every target. In the right panel with no thermal emission, we find that while the majority of targets are still optimized at 0.9 {\microns}, there are targets that are preferred at a long wavelength edge of 1.96 {\microns} (bandpass center of 1.78 {\microns}). 

\begin{figure*}
    \centering
    \includegraphics[width=0.75\linewidth]{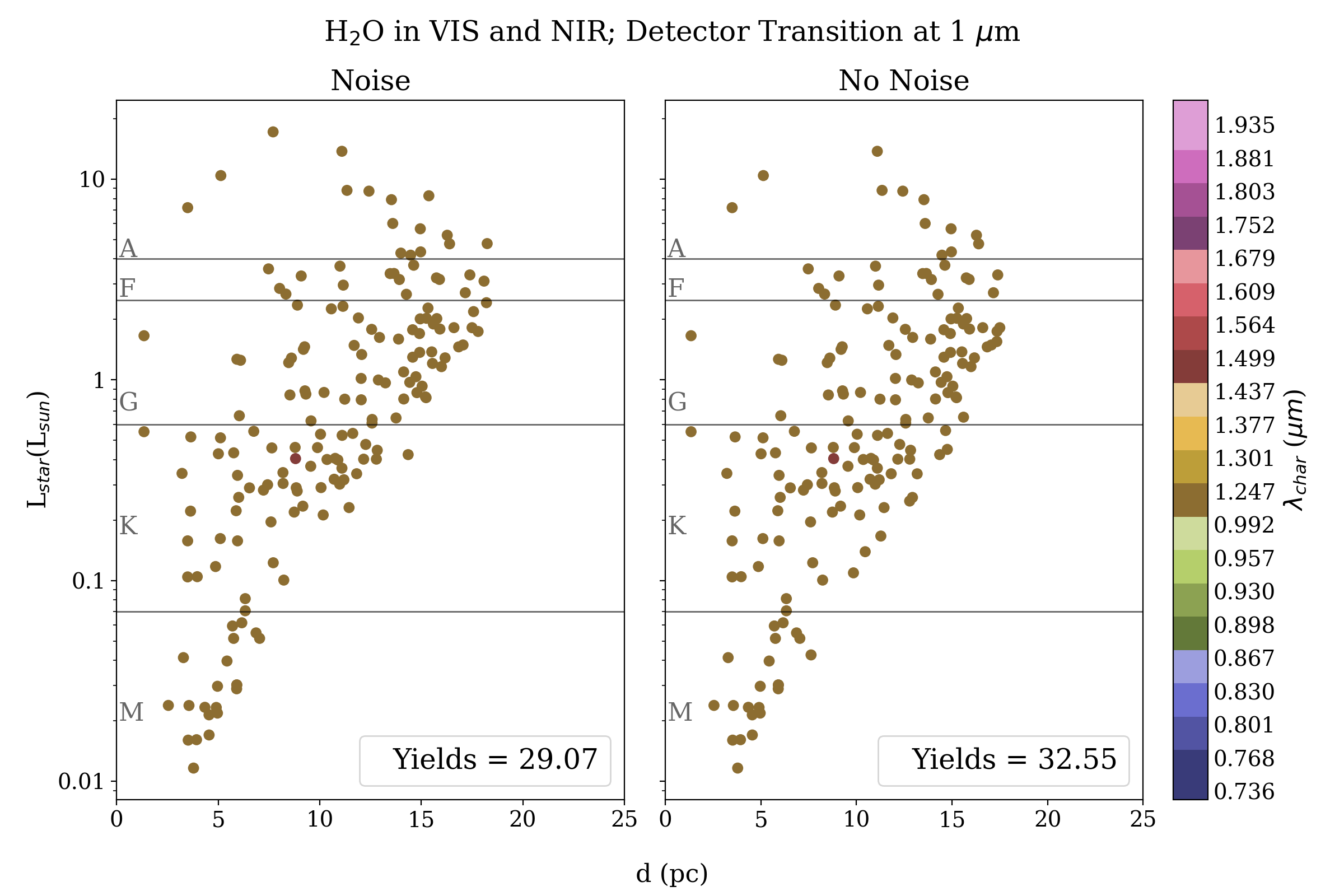}
    \caption{All plot facets identical to Figure~\ref{fig:ayo_h2o_vis_nir_1000_temp}, with full detector noise in the left panel and no detector noise in the right. Thermal emission is included in both panels. \textbf{Detector noise has a higher impact on total yields than thermal emission.}}
    \label{fig:ayo_h2o_vis_nir_1000_noise}
\end{figure*}

We present Figure~\ref{fig:ayo_h2o_vis_nir_1000_noise} to summarize the results across the VIS and NIR with and without detector noise. Thermal emission is included in both cases. We find that, conversely to the prior results, the wavelength optimization selection is identical between panels, but the yields increase when detector noise is removed, with a $\sim$3\% increase between the left panel (detector noise present) and the right panel (no detector noise). We find that, as expected, detector noise and thermal emission impact the calculated yields and preferred wavelength for optimization. While detector noise is the main driver for the total calculated yield, the presence of thermal emission drives the optimal wavelength selected for observation. This effect is most pronounced in the NIR, as thermal emission becomes a much more impactful parameter at longer wavelengths. For the most realistic calculated yields, we will assume the base detector noise case detailed in Table~\ref{tab:missionparams} for all remaining results, along with a temperature of 290K, for the remainder of this manuscript. Further work can illuminate how the intricacies of varying detectors can or will influence the calculated yields.

\subsubsection{Characterization of \ce{H2O} \& \ce{O2}}

\begin{figure*}
    \centering
    \includegraphics[width=0.75\linewidth]{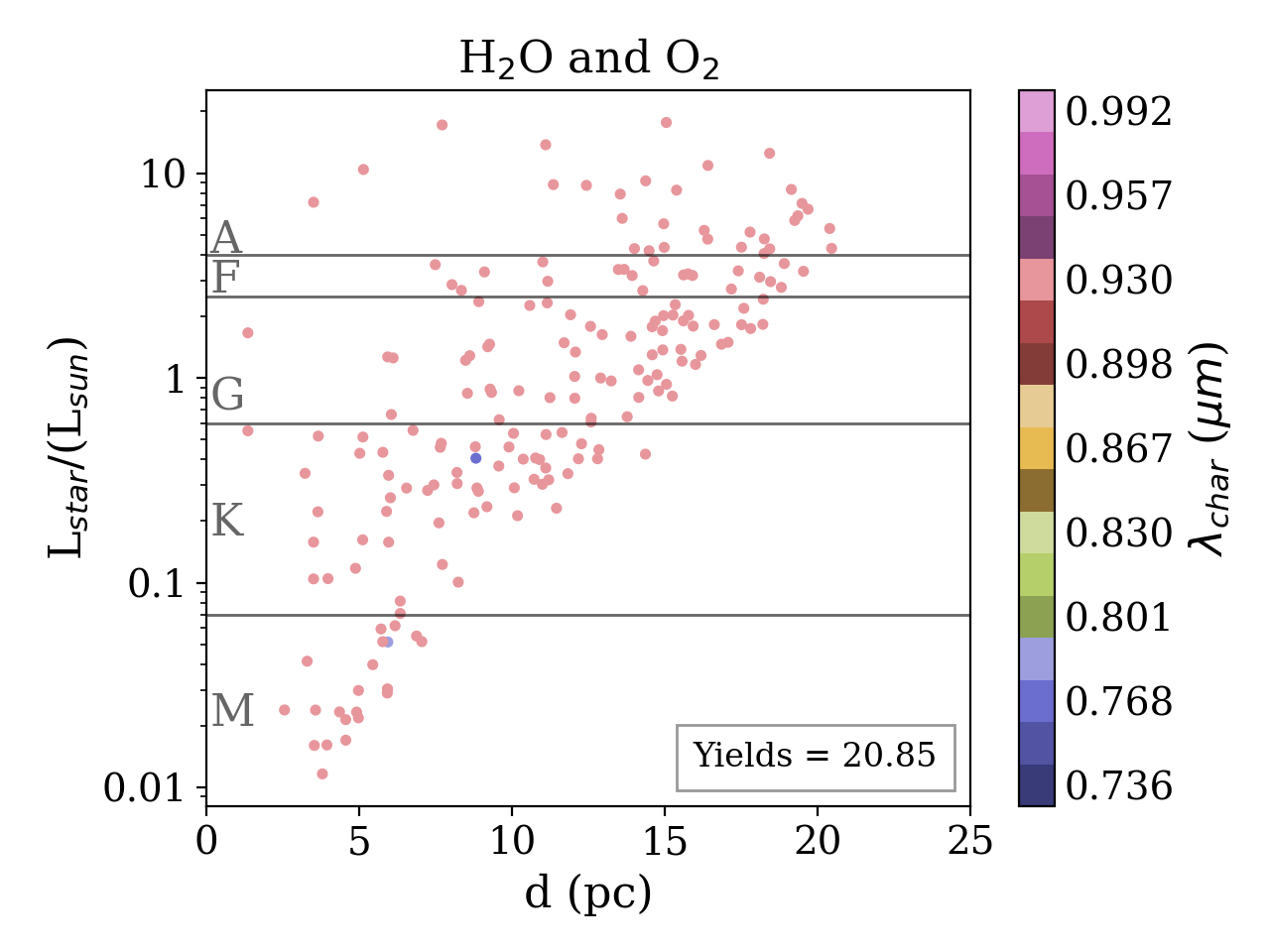}
    \caption{All plot facets identical to Figure~\ref{fig:ayo_h2o_vis_nir_1000_temp}. Simultaneous characterization of \ce{H2O} and \ce{O2}. Detector noise and thermal emission are included. \textbf{Shorter wavelengths are preferred for dual detection, but result in lower yields.}}
    \label{fig:ayo_h2o_o2}
\end{figure*}

We present the resultant yields for simultaneous detection of \ce{H2O} and \ce{O2} in the VIS in Figure~\ref{fig:ayo_h2o_o2}. We find that, compared to the \ce{H2O} only approach seen in Figure~\ref{fig:ayo_h2o_vis_nir_1000_temp}, there are significantly less calculated yields - 20.9, which is a 28.3\% decrease from the 29.1 calculated yield in the thermal emission and detector noise scenario presented in Figure~\ref{fig:ayo_h2o_vis_nir_1000_temp}. However, the dual characterization of \ce{H2O} and \ce{O2} can provide immense insight into the likely biological processes happening on the planet's surface, as a detectable level of \ce{O2} must be quite high \cite{latouf24a}. Additionally, a shorter wavelength is preferred than the \ce{H2O} only option - at a preferred 0.911 {\microns} long wavelength edge (0.83 {\microns} bandpass center), we can observe more distant and lower mass stars, and limit more astrophysical noise than at the longer wavelength 0.9 {\microns} preferred by \ce{H2O} only option. 

\subsubsection{Characterization of \ce{CH4}$/$\ce{CO2}}

We present yield calculations of possible \ce{CH4} and \ce{CO2} detection and characterization. For simplicity in this manuscript, we select the lowest abundance of \ce{CH4} and \ce{CO2} at which strong detection is possible (1.65$\times10^{-3}$ and 1$\times10^{-1}$, respectively)\cite{barbie3, hagee25}, in order to avoid over-saturated regions (such as the effect of \ce{CH4} at Archean levels \cite{barbie3}), and provide a more realistic estimation of planetary yields. This also results in a more constrained possible wavelength region for detections centered on the molecular feature, while also requiring higher SNRs for strong detection, thus investigating the preferred wavelength under more conservative conditions. 


\begin{figure*}
    \centering
    \includegraphics[width=0.75\linewidth]{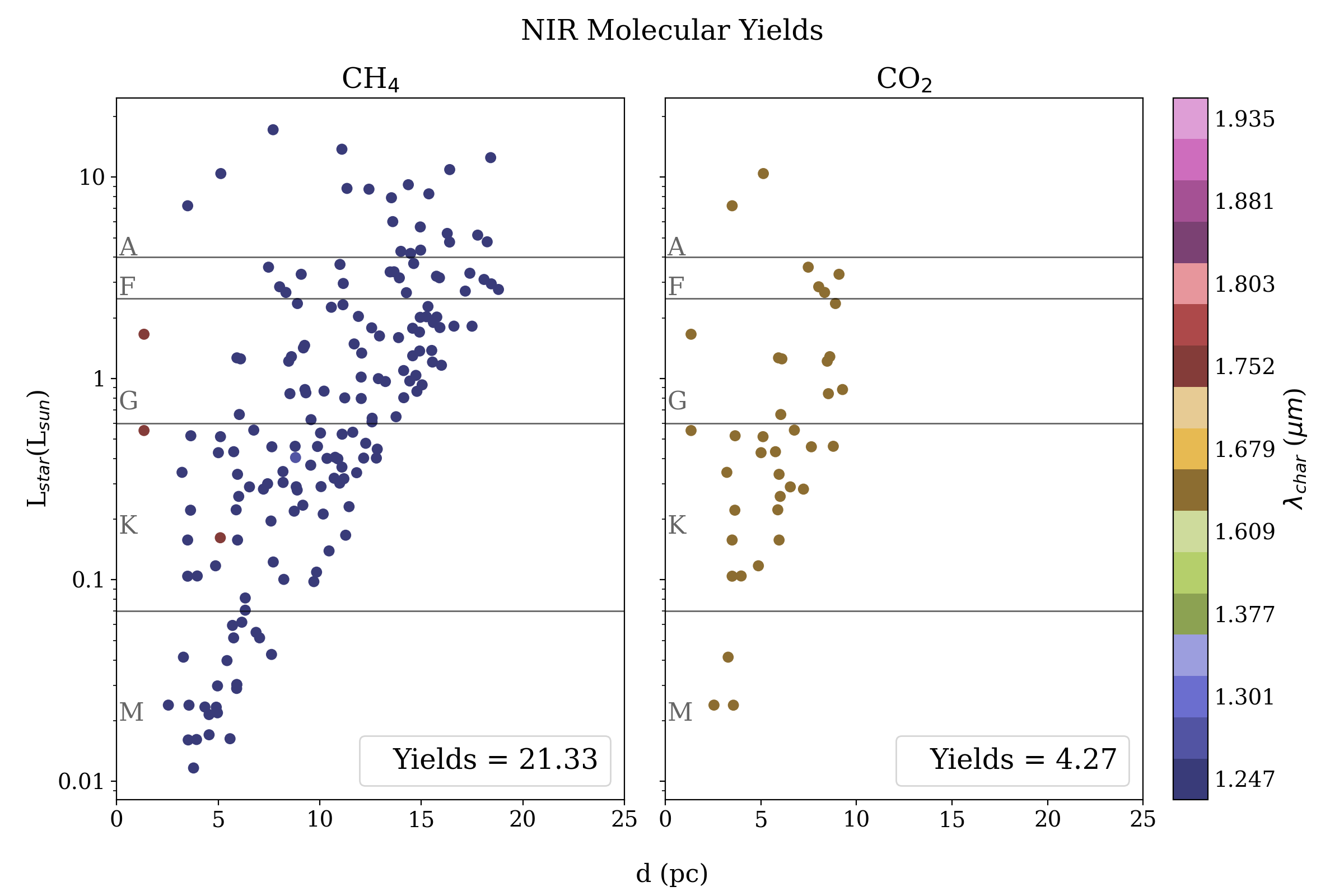}
    \caption{All plot facets identical to Figure~\ref{fig:ayo_h2o_vis_nir_1000_temp}. Detector noise is fully included. \textit{Left Panel:} exoEarth yields for the detection and characterization of \ce{CH4}. \textit{Right Panel:} exoEarth yields for the detection and characterization of \ce{CO2}. \textbf{\ce{CH4} results in significantly higher yields than \ce{CO2}, with \ce{CO2} preferring longer wavelengths in the NIR than \ce{CH4}.}}
    \label{fig:ayo_nir}
\end{figure*}

We present our results in Figure~\ref{fig:ayo_nir}. We present the yields for \ce{CH4} in the left panel, and \ce{CO2} in the right panel. Beginning with the left panel, we find that \ce{CH4} is preferred at one main long wavelength edge - 1.25 {\microns} (bandpass center at 1.1 {\microns}), as a function of luminosity and distance. The habitable-zone planets around lower-mass stars are more difficult to observe due to their closer proximity to the star and thus are more limited by the lower range of the inner working angle (IWA), thus shorter wavelengths are preferred. This showcases the utility in a star-by-star analysis. The resultant yields are 21.3, which is a 26.6\% decrease in yields compared to searching for \ce{H2O} alone, and a 2.3\% increase in yields compared to searching for \ce{H2O} and \ce{O2} simultaneously. Thus, searching for \ce{CH4} in the NIR is a comparable survey strategy to dual detecting \ce{H2O} and \ce{O2} in the VIS, depending on the desired scientific outcome. 

In the right panel of Figure~\ref{fig:ayo_nir}, we present the yields for \ce{CO2}. We find that \ce{CO2} is preferred at the 1.65 {\microns} long wavelength edge (1.46 {\microns} bandpass center). However, notably, no targets were selected further than $\sim$12 pc away, and the yields were significantly lessened compared to other survey strategies at $\sim$4.3, with a decrease of 85.3\%, 79.5\%, and 80\% compared to searching for \ce{H2O}, \ce{H2O} and \ce{O2}, and \ce{CH4}, respectively. Thus, searching for \ce{CO2} is not a comparable strategy to any of the other strategies presented.

\section{Discussion \& Future Work}
\label{sec:discuss}

\begin{table}[h!]
\centering
\resizebox{0.3\columnwidth}{!}{%
\begin{threeparttable}
\caption{AYO Yield Results with Detector Noise and Thermal Emission}
\label{tab:yields}
\begin{tabular}{c|c}
\hline
\hline
\textbf{Molecule} & \textbf{Calculated Yield}  \\
\hline
\ce{H2O} & \textbf{29.1} \\
\ce{H2O} \& \ce{O2} & \textbf{20.9} \\
\ce{CH4} & \textbf{21.3} \\
\ce{CO2} & \textbf{4.3} \\
\hline
\hline
\end{tabular}
\end{threeparttable}}
\end{table}
The detection and characterization of \ce{H2O} can vary in efficacy and exoEarth yields through wavelength, and depending on the software used. Figure~\ref{fig:h2o_ayo_exosims} presents the yields per wavelength using both AYO and EXOSIMS to characterize \ce{H2O} in the VIS. We find that there is consensus in observing the 0.9 {\microns} \ce{H2O} feature, although the calculated yields vary between AYO and EXOSIMS. Additionally, the calculated yield pattern is also mirrored between AYO and EXOSIMS, with the longer wavelengths resulting in the highest calculated yields. In Figure~\ref{fig:h2o_o2_ayo_exosims}, investigating the dual detection of \ce{H2O} and \ce{O2} in the VIS with AYO and EXOSIMS, we find that the two highest yields are calculated at the same two wavelengths. AYO finds a slightly preferred wavelength of observation centering on 0.83 {\microns} (long wavelength edge of 0.911 {\microns}), and EXOSIMS also finds a preferred wavelength at 0.83 {\microns}. All of the EXOSIMS results are additionally within an average standard deviation of 2.51, leading to inconclusive results. We note that while the relative trends in preferred wavelength are consistent, the raw yield magnitudes differ. Resolving these quantitative discrepancies requires a dedicated cross-validation campaign between AYO and EXOSIMS, which is beyond the scope of this work. Consequently, we focus here on the agreement in relative behavior to drive instrument design requirements. Presently, AYO additionally has the capability to optimize wavelength on a target-by-target basis for specific characterization scenarios, thus avoiding the need to calculate yields for multiple wavelengths separately. 

A high-level summary of wavelength-optimized AYO yield results is presented in Table~\ref{tab:yields}. In addition to the LUVOIR final report, other works such as Refs.~\citenum{feng18, latouf23, stark24a} have investigated the preferred wavelength for \ce{H2O} detection as a first step. However, each of these works were conducted only in the VIS regime. With \ce{H2O} as a highly absorbent feature across multiple wavelengths, Figure~\ref{fig:ayo_h2o_vis_nir_1000_temp} illustrates that the preferred wavelength for characterization of \ce{H2O} remains the 0.9 {\microns} feature. This is crucial when looking into instrument design, as we can determine that \ce{H2O} is not a driver for the NIR detector development. When investigating the impact of thermal emission, in the right panel of Figure~\ref{fig:ayo_h2o_vis_nir_1000_temp}, we find that while the 0.9 {\microns} feature is still preferred for most targets, the feature near 1.78 {\microns} is optimal for several targets when there is no thermal emission. When thermal emission is included, as in the left panel of Figure~\ref{fig:ayo_h2o_vis_nir_1000_temp}, we find that all targets except one are preferred at 0.9 {\microns}. The calculated yield change is minimal, at less than 0.05\%, thus thermal emission is a major driver in optimal observational wavelength, but not total yields. This is additionally critical to understand when we see that detector noise influences the total resultant yields, and not the observational wavelength, as seen in Figure~\ref{fig:ayo_h2o_vis_nir_1000_noise}. With the detector noise scenario detailed in Table~\ref{tab:missionparams}, the total calculated yield is 29.1, compared to 32.6 assuming no detector noise, resulting in a 12\% increase in yields with no detector noise. However, the preferred wavelength for observation remains at 0.9 {\microns}. Thus, we can conclude that \ce{H2O}, in almost every scenario, is most productively observed at 0.9 {\microns} and to increase yields, detector development that minimizes noise is most critical.   

As \ce{H2O} is preferred to be observed at the 0.9 {\microns} feature (corresponding to a long wavelength edge of $\sim$1 {\microns}) rather than deeper into the NIR, the natural next step is to investigate whether shorter wavelengths, where dual molecular detection of \ce{H2O} and \ce{O2} in the VIS is possible, result in promising yields. In Figure~\ref{fig:ayo_h2o_o2}, we find that the yields decrease by 28\% from those presented in the \ce{H2O} only case (assuming inclusion of thermal emission and detector noise), however with the notable difference that the resultant outcomes are planets with both \ce{H2O} and \ce{O2} characterized at modern Earth levels. This is under the large caveat that \ce{O2} is at undetectable abundances at broadband SNRs of 20 or below for most of Earth's history\cite{latouf24a}. Modern levels of \ce{O2} are detectable concurrently with \ce{H2O} \cite{latouf23}, and abundances as low as 50\% present atmospheric levels (PALs) are detectable either individually or dually with \ce{H2O} at high SNRs depending on the wavelength of observation; but for lower than 50\% PAL \ce{O2}, an SNR higher than 20 will be required. This would make identification of a Proterozoic Earth difficult, as the \ce{O2} levels are believed to be approximately 0.1\% to 1\% PAL \cite{planavsky14}. This can be considered a benefit of this strategy -- by aiming to detect \ce{H2O} and \ce{O2} simultaneously, if \ce{O2} is not detected at the expected SNR but \ce{H2O} is detected, we can begin to draw conclusions on the potential corresponding Earth epoch. However, this is very specific to an Earth-like atmosphere -- many other exo-atmospheres can present \ce{H2O} and not \ce{O2}. Thus, although the visible regime is hugely beneficial, understanding the yields of planets with molecules in the NIR are critical for providing context. 

In Figure~\ref{fig:ayo_nir} we investigate the resultant yields for characterizing \ce{CH4} (left panel) and \ce{CO2} (right panel) in the NIR. We find that searching for \ce{CO2} as a first order molecule is not an efficient observational strategy, with the lowest resultant yields compared to other strategies investigated at 4.3. Additionally, \ce{CO2} is not detectable at targets further than 12 pc, cutting the range of distant targets possible for observation. The preferred central wavelength for observation was 1.5 {\microns}, which would approach the proposed long wavelength cut-off for the coronagraph \cite{jktlivingworlds, hagee25} and thus incur additional error. Searching for \ce{CH4}, however, results in yields of 21.3. Although that is a $\sim$26\% decrease in yields from the search for \ce{H2O} presented in Figure~\ref{fig:ayo_h2o_vis_nir_1000_temp}, it is an increase of $\sim$2\% from the search for \ce{H2O} and \ce{O2} presented in Figure~\ref{fig:ayo_h2o_o2}. If we assume an observational decision tree that begins with detection and characterization of \ce{H2O}, the following step could thus be either the search for \ce{O2} in the VIS or \ce{CH4} in the NIR without a critical decrease in exoEarth yields and informational content. The preferred wavelength for observation of \ce{CH4} is at a central wavelength of 1.1 {\microns}. The information content of dimmer stars tends to peak in the red-optical and very near infrared, especially for M-dwarfs, thus the shorter wavelength for preferred observation is likely selected to account for higher informational content \cite{reiners2018,reiners2010}. Prior work by \citenum{hagee25} has shown that in the NIR, \ce{CO2}, \ce{CH4} and \ce{H2O} are degenerate with each other, resulting in detection influence as a function of abundance (i.e. as one molecule becomes more abundant and detectable, the detectability of another molecule decreases). Further work is required to understand the possible yields for simultaneous characterizations of \ce{CO2}, \ce{CH4} and/or \ce{H2O} in the NIR region, with careful consideration of the possible degeneracies. 

Many of the survey strategies presented here were not able to be investigated with yields at this time, however, the development of the survey strategies the yield codes can be updated accordingly. In order to investigate these remaining strategies, changes to existing yield calculator codes are necessary. EXOSIMS operates with a scheduler module; to change the criteria for the order of observations, a new scheduler module will need to be created and inserted. This new scheduler must be able to begin with, for instance, molecular characterization and then move to orbit measurements, rather than the currently adopted detection, orbit, and characterization pipeline. Currently, work is underway to implement the ability to observe in series or parallel with pre-determined wavelength bands and, in the long term, to develop a scheduler that will use an appropriate wavelength band per observation stepping through a decision tree framework. 

AYO currently does not have the ability to conduct multi-tiered observational yields, wherein each step has an intrinsic yield and each pathway has a total resultant yield. In future work, AYO will be adapted to enable multi-tiered decision frameworks for yield calculation. Ideally, up to four subsequent decisions will be possible; however it is computationally expensive to demand that each decision is predicated on the past one. By instead establishing multiple pathways and enabling AYO to step through said pathways, yields for different observational strategies will be possible. Future work will conduct multi-step yield analyses of observation strategies, as well as the implications of survey strategy for the hardware configurations of HWO \cite{latouf26}.

Very recently, a proof-of-concept study has been carried out to evaluate the impact of multi-bandpass photometry, i.e. observing in multiple bandpasses rather than just VIS, during the discovery phase. This work suggests that color-based photometry might allow for a qualitative differentiation between Earth-like and Neptune-like planets, thus providing additional information from the detection phase to inform subsequent follow-up characterization \cite{alei2025phot}. Further studies could be conducted to understand if precursor information, whether indirect (e.g., extreme precision radial velocity mass measurements, $\msini$, orbital parameters, or transit data) or direct (e.g., ground detection of promising planet candidates orbiting M-dwarfs with the future Extremely Large Telescopes (ELTs)), could change the observational strategy. The reliability of biosignature interpretation also critically depends on mitigating degeneracies. For instance, abiotic \ce{O2}/\ce{O3} production can mimic biological oxygenation, while high-altitude clouds and aerosols may mask \ce{H2O} and other features. Each strategy should therefore be evaluated in terms of the vulnerability to abiotic false positives, the risk of false negatives, possible additional molecular discriminators, and the complementarity of modeling and observables, such as the context of all the molecules observed taken together, while also keeping in mind the risk and feasibility. 

\section{Conclusions}
\label{sec:conc}

We find that observing \ce{H2O} following planetary detection and orbit measurement is well-studied and replicable, with a preferred characterization wavelength of 0.9 {\microns} in both previous VIS-only simulations and when given both VIS and NIR options. Thermal emission is the main driver of preferred wavelength for characterization, with more targets selecting long wavelengths when thermal emission is not considered, but the detector noise is the main driver in decreasing or increasing the resultant calculated yields. Dual characterization of \ce{H2O} and \ce{O2} results in lower yields, but potentially more informational content than searching for a single molecule (\ce{H2O}). \ce{CH4} and \ce{CO2} result in lower yields individually, with future work necessary to understand the known degeneracies between \ce{H2O}, \ce{CH4}, and \ce{CO2} and the influence on yields, however the yields for \ce{CH4} are comparable to the yields searching for \ce{H2O} and \ce{O2} simultaneously. Depending on the desired molecule of interest, both paths can be utilized in the search for Earth-like worlds. Many survey strategies that follow more unorthodox observational methods have yet to be investigated, but current work is underway to understand the most efficient observational decision tree per planet and stellar type. 



\subsection*{Disclosures}
The authors declare that there are no financial interests, commercial affiliations, or other potential conflicts of interest that could have influenced the objectivity of this research or the writing of this paper” is included in a Disclosures section of the manuscript.

\subsection* {Code, Data, and Materials Availability} 
NASA regulations govern the release of source code, including what can be released and how it is made available. Readers should contact the corresponding author if they would like copies of the visualization software or data produced for this study.

\subsection* {Acknowledgments}
N. L. and E.A. gratefully acknowledges financial support by an appointment to the NASA Postdoctoral Program (NPP) at the NASA Goddard Space Flight Center, administered by Oak Ridge Associated Universities under contract with NASA. ES acknowledges support from grants 80NSSC23K0039, 80NSSC23K1399, and 80NSSC23K1398. The views and conclusions contained in this document are those of the authors and should not be interpreted as representing the official policies, either expressed or implied, of the National Aeronautics and Space Administration (NASA) or the U.S. Government. The U.S. Government is authorized to reproduce and distribute reprints for Government purposes notwithstanding any copyright notation herein. The authors would also like to thank the Habitable Worlds Observatory (HWO) START and TAG teams for their efforts, as well as the numerous working groups and task groups nested underneath the START and TAG. 



\bibliography{main}   

\begin{thebibliography}{10}

\bibitem{gaplist}
{NASA JPL}, ``Exoplanet exploration program technology gap list,''  (2024).
\newblock Accessed: 2025-09-03.

\bibitem{luvoir}
{The LUVOIR Team}, ``{The LUVOIR Mission Concept Study Final Report},'' {\em arXiv e-prints} , arXiv:1912.06219  (2019).

\bibitem{habex}
B.~S. {Gaudi}, S.~{Seager}, B.~{Mennesson}, {\em et~al.}, ``{The Habitable Exoplanet Observatory (HabEx) Mission Concept Study Final Report},'' {\em arXiv e-prints} , arXiv:2001.06683  (2020).

\bibitem{livingworldsscdd}
G.~{Arney}, N.~{Parenteau}, N.~{Hinkel}, {\em et~al.}, ``{Habitable Worlds Observatory (HWO): Living Worlds Community Working Group: The Search for Life on Potentially Habitable Exoplanets},'' {\em arXiv e-prints} , arXiv:2601.09766  (2026).

\bibitem{jktlivingworlds}
J.~{Krissansen-Totton}, A.~G. {Ulses}, M.~{Frissell}, {\em et~al.}, ``{Wavelength Requirements for Life Detection via Reflected Light Spectroscopy of Rocky Exoplanets},'' {\em arXiv e-prints} , arXiv:2507.14771  (2025).

\bibitem{stark19}
C.~C. {Stark}, R.~{Belikov}, M.~R. {Bolcar}, {\em et~al.}, ``{ExoEarth yield landscape for future direct imaging space telescopes},'' {\em Journal of Astronomical Telescopes, Instruments, and Systems} {\bf 5}, 024009  (2019).

\bibitem{stark24b}
C.~C. {Stark}, B.~{Mennesson}, S.~{Bryson}, {\em et~al.}, ``{{Paths to Robust Exoplanet Science Yield Margin for the Habitable Worlds Observatory}},'' {\em arXiv e-prints} , arXiv:2405.19418  (2024).

\bibitem{savransky16}
D.~{Savransky} and D.~{Garrett}, ``{WFIRST-AFTA coronagraph science yield modeling with EXOSIMS},'' {\em Journal of Astronomical Telescopes, Instruments, and Systems} {\bf 2}, 011006  (2016).

\bibitem{savransky2017exosims}
D.~{Savransky}, C.~{Delacroix}, and D.~{Garrett}, ``{EXOSIMS: Exoplanet Open-Source Imaging Mission Simulator}.'' Astrophysics Source Code Library  (2017).

\bibitem{brown2004}
R.~A. {Brown}, ``{Obscurational Completeness},'' {\em \apj} {\bf 607}, 1003--1013  (2004).

\bibitem{tuchow24}
N.~W. {Tuchow}, C.~C. {Stark}, and E.~{Mamajek}, ``{HPIC: The Habitable Worlds Observatory Preliminary Input Catalog},'' {\em \aj} {\bf 167}, 139  (2024).

\bibitem{latoufresolving}
N.~{Latouf}, C.~{Stark}, A.~{Mandell}, {\em et~al.}, ``{Determining the Detectability of H2O with Photometric Observations using Bayesian Analysis for Remote Biosignature Identification on exoEarths (BARBIE)},'' {\em arXiv e-prints} , arXiv:2512.07620  (2025).

\bibitem{savransky10}
D.~{Savransky}, N.~J. {Kasdin}, and E.~{Cady}, ``{Analyzing the Designs of Planet-Finding Missions},'' {\em \pasp} {\bf 122}, 401  (2010).

\bibitem{stark24a}
C.~C. {Stark}, N.~{Latouf}, A.~M. {Mandell}, {\em et~al.}, ``{Optimized bandpasses for the Habitable Worlds Observatory's exoEarth survey},'' {\em Journal of Astronomical Telescopes, Instruments, and Systems} {\bf 10}, 014005  (2024).

\bibitem{stark25arxiv}
C.~C. {Stark}, S.~{Steiger}, A.~{Tokadjian}, {\em et~al.}, ``{Cross-Model Validation of Coronagraphic Exposure Time Calculators for the Habitable Worlds Observatory: A Report from the Exoplanet Science Yield sub-Working Group},'' {\em arXiv e-prints} , arXiv:2502.18556  (2025).

\bibitem{bebek}
C.~J. Bebek, J.~H. Emes, D.~E. Groom, {\em et~al.}, ``{Status of the CCD development for the Dark Energy Spectroscopic Instrument},'' {\em JINST} {\bf 12}, C04018----C04018  (2017).

\bibitem{teledyne}
{Teledyne-e2v}, ``{Scientific CCD Image Sensors},''  (2024).
\newblock Accessed: 2024-06-06.

\bibitem{romanemccd}
{IPAC}, ``Spacecraft and instrument parameters,''  (2025).
\newblock Accessed: 2025-08-12.

\bibitem{latouf23}
N.~{Latouf}, A.~M. {Mandell}, G.~L. {Villanueva}, {\em et~al.}, ``{Bayesian Analysis for Remote Biosignature Identification on exoEarths (BARBIE). I. Using Grid-based Nested Sampling in Coronagraphy Observation Simulations for H2O},'' {\em \aj} {\bf 166}, 129  (2023).

\bibitem{latouf24a}
N.~{Latouf}, A.~M. {Mandell}, G.~L. {Villanueva}, {\em et~al.}, ``{Bayesian Analysis for Remote Biosignature Identification on exoEarths (BARBIE). II. Using Grid-based Nested Sampling in Coronagraphy Observation Simulations for O2 and O3},'' {\em \aj} {\bf 167}, 27  (2024).

\bibitem{barbie3}
N.~{Latouf}, M.~D. {Himes}, A.~M. {Mandell}, {\em et~al.}, ``{BARBIE. Bayesian Analysis for Remote Biosignature Identification on exoEarths. III. Introducing the KEN},'' {\em \aj} {\bf 169}, 50  (2025).

\bibitem{Spohn2025}
C.~Spohn, S.~Steiger, and A.~R. Howe, ``yieldplotlib: A unified library for exoplanet yield code visualizations,'' {\em Journal of Open Source Software} {\bf 10}(116), 9401  (2025).

\bibitem{dulzJointRadial2020}
S.~D. Dulz, P.~Plavchan, J.~R. Crepp, {\em et~al.}, ``Joint {{Radial Velocity}} and {{Direct Imaging Planet Yield Calculations}}. {{I}}. {{Self-consistent Planet Populations}},'' {\em The Astrophysical Journal} {\bf 893}, 122  (2020).

\bibitem{spohn2026}
C.~Spohn, C.~C. Stark, D.~Savransky, {\em et~al.}, ``Understanding {{HWO}}'s field of regard and characterization requirement trade space with a dynamic observation scheduling algorithm,'' {\em Journal of Astronomical Telescopes, Instruments, and Systems} {\bf 12}, 041010  (2026).

\bibitem{horning19}
A.~{Horning}, R.~{Morgan}, and E.~{Nielson}, ``{Minimum number of observations for exoplanet orbit determination},'' in {\em Society of Photo-Optical Instrumentation Engineers (SPIE) Conference Series},  {\em Society of Photo-Optical Instrumentation Engineers (SPIE) Conference Series} {\bf 11117}, 111171C  (2019).

\bibitem{blunt17}
S.~{Blunt}, E.~L. {Nielsen}, R.~J. {De Rosa}, {\em et~al.}, ``{Orbits for the Impatient: A Bayesian Rejection-sampling Method for Quickly Fitting the Orbits of Long-period Exoplanets},'' {\em \aj} {\bf 153}, 229  (2017).

\bibitem{domagalgoldman14}
S.~D. {Domagal-Goldman}, A.~{Segura}, M.~W. {Claire}, {\em et~al.}, ``{Abiotic Ozone and Oxygen in Atmospheres Similar to Prebiotic Earth},'' {\em \apj} {\bf 792}, 90  (2014).

\bibitem{meadows17}
V.~S. {Meadows}, ``{Reflections on O$_{2}$ as a Biosignature in Exoplanetary Atmospheres},'' {\em Astrobiology} {\bf 17}, 1022--1052  (2017).

\bibitem{savranskyo3}
D.~{Savransky}, D.~N. {Spergel}, N.~J. {Kasdin}, {\em et~al.}, ``{Occulting ozone observatory science overview},'' in {\em Space Telescopes and Instrumentation 2010: Optical, Infrared, and Millimeter Wave},  J.~M. {Oschmann}, Jr., M.~C. {Clampin}, and H.~A. {MacEwen}, Eds., {\em Society of Photo-Optical Instrumentation Engineers (SPIE) Conference Series} {\bf 7731}, 77312H  (2010).

\bibitem{o3mission}
D.~{Lisman} and E.~W. {Schwieterman}, ``{The Occulting Ozone Observatory (O3) Mission},'' in {\em Bulletin of the American Astronomical Society},   {\bf 51}, 217  (2019).

\bibitem{starkbenchmark}
C.~C. {Stark}, S.~{Steiger}, A.~{Tokadjian}, {\em et~al.}, ``{Cross-Model Validation of Coronagraphic Exposure Time Calculators for the Habitable Worlds Observatory: A Report from the Exoplanet Science Yield sub-Working Group},'' {\em arXiv e-prints} , arXiv:2502.18556  (2025).

\bibitem{feng18}
Y.~K. {Feng}, T.~D. {Robinson}, J.~J. {Fortney}, {\em et~al.}, ``{Characterizing Earth Analogs in Reflected Light: Atmospheric Retrieval Studies for Future Space Telescopes},'' {\em \aj} {\bf 155}, 200  (2018).

\bibitem{hagee25}
C.~{Hagee}, N.~{Latouf}, A.~M. {Mandell}, {\em et~al.}, ``Bayesian analysis for remote biosignature identification on exoearths (barbie) iv: Analyzing \ce{CO2} detections in the near-ir to determine the long-wavelength cut-off for the habitable worlds observatory coronagraph.''  (submitted).

\bibitem{planavsky14}
N.~J. {Planavsky}, C.~T. {Reinhard}, X.~{Wang}, {\em et~al.}, ``{Low Mid-Proterozoic atmospheric oxygen levels and the delayed rise of animals},'' {\em Science} {\bf 346}, 635--638  (2014).

\bibitem{reiners2018}
A.~{Reiners}, M.~{Zechmeister}, J.~A. {Caballero}, {\em et~al.}, ``{The CARMENES search for exoplanets around M dwarfs. High-resolution optical and near-infrared spectroscopy of 324 survey stars},'' {\em \aap} {\bf 612}, A49  (2018).

\bibitem{reiners2010}
A.~{Reiners}, J.~L. {Bean}, K.~F. {Huber}, {\em et~al.}, ``{Detecting Planets Around Very Low Mass Stars with the Radial Velocity Method},'' {\em \apj} {\bf 710}, 432--443  (2010).

\bibitem{latouf26}
N.~{Latouf}, , A.~{Young}, {\em et~al.}, ``Utilizing bayesian analysis of remote biosignature identification on exoearths (barbie) methodology to update a decision tree observational framework.''  (in prep).

\bibitem{alei2025phot}
E.~{Alei}, A.~M. {Mandell}, M.~H. {Currie}, {\em et~al.}, ``{Multi-bandpass Photometry for Exoplanet Atmosphere Reconnaissance (MPEAR) with the Habitable Worlds Observatory (HWO) -- I. Differentiating Earth from Neptunes During Discovery},'' {\em arXiv e-prints} , arXiv:2512.05279  (2025).

\end{thebibliography}
\bibliographystyle{spiejour}   


\vspace{1ex}
\noindent Biographies and photographs of the authors are not available.

\listoffigures
\listoftables

\end{spacing}
\end{document}